\documentclass[11pt,twoside]{article}

\usepackage{asp2006}
\usepackage{epsf}
\usepackage{epsfig} 
\usepackage{lscape}
\usepackage{amssymb,amsmath}

\begin{document}
\title{Particle Dark Matter: Status and Searches} 
\author{Pearl Sandick}  
\affil{Theory Group and Texas Cosmology Center, The University of Texas at Austin, TX 78712}   

\begin{abstract}
A brief overview is given of the phenomenology of particle dark matter and the properties of some of the most widely studied dark matter candidates.  Recent developments in direct and indirect dark matter searches are discussed.
\end{abstract}



\section{Introduction}

It has been suspected for over three quarters of a century that at least some portion of the matter in the universe must be non-luminous.  Fritz Zwicky provided the first evidence of dark matter with the finding that the peculiar velocities of galaxies at the edge of the Coma Cluster were much larger than what one would expect if all the matter in the cluster were luminous~\citep{zwicky}.  Although the idea of dark matter took some time to gain traction in the community, eventually, measurements of galactic rotation curves revealed that either Newtonian mechanics is at odds with galactic motion, or that dark matter plays a substantial role on galactic scales as well~\citep{rubin}. Over the last few decades, precision cosmological measurements and simulations of structure formation have confirmed, again and again, the cold dark matter paradigm, and we now know, thanks to measurements of the Cosmic Microwave Background (CMB), baryon acoustic oscillations, and Type-Ia supernovae, the abundance of dark matter at the few percent level~\citep{wmap}:
\begin{equation}
\Omega_{CDM}=0.228\pm0.013.
\end{equation}

Verifications of the hypothesis that gravitating non-luminous matter is the cause of the observed galactic and cluster dynamics have been possible recently with the combination of gravitational lensing measurements and X-ray observations.  The most noted case is that of the Bullet Cluster~\citep{bullet}, in which the optical and X-ray components of two colliding galaxy clusters are clearly separated.  Gravitational lensing is used to determine the mass distribution, and it was shown that the mass is concentrated near the luminous objects, which interact only gravitationally in the collision.  The hot gas, visible in X-rays, however, interacts electromagnetically as well.  Therefore, the hot gas, which makes up most of the baryonic matter in the cluster, is not the main source of gravitating matter, and a dark matter component is required.  This example is particularly important in that theories of modified gravitational interactions 
can not explain the Bullet Cluster scenario without also invoking a cold dark matter component. Theories of Modified Newtonian Dynamics are therefore disfavored~\citep{mondout}.

In addition to the wealth of evidence that non-luminous matter makes up a substantial portion of the energy budget of the universe, the success of the theory of Big Bang Nucleosynthesis and its agreement with the baryon density from CMB observations fixes the amount of baryonic matter in the universe, pointing to the conclusion that cold dark matter does not consist of Standard Model particles\footnote{Neutrinos were long considered to be a dark matter condidate in the Standard Model, however, due to the smallness of neutrino masses, the relic density of neutrinos would be far too small to account for the dark matter in the universe.  Their abundance is also strongly constrained by CMB anisotropies and large scale structure data.}.
There are, however, a variety of dark matter candidates in well-motivated theories of particle physics beyond the Standard Model.

In order to determine the nature of particle dark matter, we must first detect it: either directly, through dark matter scattering on nuclei in terrestrial detectors; indirectly, through observations of its annihilation or decay products; or by producing it in colliders.  Here, I briefly discuss a few of the main particle candidates for cold dark matter and the direct and indirect detection experiments that may help to elucidate its nature.

\section{Candidates}

Independent of the dark matter question, the Standard Model of particle physics is an incomplete theory.  It describes the particles and their interactions remarkably well at energies explored to date at particle colliders, but there are some very serious theoretical reasons to expect that the Standard Model is only part of a larger theoretical framework.  Many such frameworks, or extensions to the Standard Model, have been proposed.  A remarkable fraction of these contain one or more dark matter candidates, often with masses in the 100 GeV to TeV range.  Particles with GeV to TeV masses and weak-scale interaction cross sections are known as Weakly Interacting Massive Particles (WIMPs).

By no means are all dark matter candidates WIMPs.  However, WIMP dark matter is compelling for two main reasons.  First, WIMPs that were in thermal equilibrium in the early universe and annihilated with each other with a roughly weak-scale cross section would have frozen out as the universe expanded and cooled to obtain a relic density today that is within an order of magnitude of the measured dark matter density. This coincidence that a particle with roughly weak scale mass and annihilation cross section would have a relic density in the ballpark of that expected from cosmological observations is known as the WIMP Miracle.
Second, as mentioned above, because new particle physics is expected just above the weak scale, WIMP dark matter candidates are present in many of the favored extensions of the Standard Model.

Two of the main theories of physics beyond the Standard Model that contain WIMP dark matter are supersymmetry and models with Universal Extra Dimensions (UED).  In supersymmetric theories, additional symmetries beyond those of the Standard Model indicate that every Standard Model particle has a supersymmetric partner with spin different by a half integer.  Standard Model fermions, such as electrons, have supersymmetric partners with spin 0, while Standard Model gauge bosons have supersymmetric fermionic partners.  The imposition of R-parity, which distinguishes supersymmetric particles from their Standard Model counterparts and prevents such undesired effects as proton decay, results in the stability of the Lightest Supersymmetric Particle (LSP).  If neutral, the LSP is a dark matter candidate.  Similarly, in models with UED, where the Standard Model particles propagate in one or more compact extra dimensions, the lightest Kaluza-Klein Particle (LKP) is a particle candidate for cold dark matter, similarly stabilized by the introduction of KK-parity.  The LKP is often the first excitation of the hypercharge gauge boson, denoted $B^{(1)}$.

Supersymmetry provides us with a host of potential dark matter candidates. The LSP may be a neutralino, $\widetilde{\chi}$, which is an admixture of supersymmetric partners of the neutral gauge bosons and CP even Higgs states.  This particle is often refered to as the ``canonical WIMP.''  Alternatively, the LSP may be the supersymmetric spin $3/2$ partner of the graviton, called the gravitino, $\widetilde{G}$, whose gravitational-strength interactions make it very difficult to discover.  The gravitino mass is related to the supersymmetry-breaking mechanism: If supersymmetry breaking is gauge-mediated, we expect a light gravitino with $m_{3/2} \lesssim$ 1 GeV~\citep{gaugeGrav}, while for gravity-mediated supersymmetry breaking, the gravitino mass would be in the same range as the masses of the other supersymmetric scalar particles, 100 GeV $\lesssim m_{3/2} \lesssim$ 1 TeV~\citep{gravGrav}. For anomaly mediation, the gravitino would be much heavier, with $m_{3/2} \gtrsim 10$ TeV~\citep{anomGrav}. Only in the former two cases might the gravitino be the LSP.  In some supersymmetric extensions, the right-handed sneutrino, the supersymmetric partner of the neutrino, is the LSP and a WIMP dark matter candidate.

Another commonly discussed dark matter candidate is the axion, which was introduced as a solution to the problem of why strong interactions conserve P and CP while the Standard Model, in general, does not. Although the axion was originally thought to be a more massive, more strongly interacting particle, the fact that they were not discovered led to upper limits on their mass and interaction strength, and eventually cosmological constraints resulted in an axion mass window of roughly $10^{-6}$ eV $\lesssim m_a \lesssim 10^{-3}$ eV~\citep{KolbTurner}.  Because of the production mechanisms, axions would be very cold, never having been in thermal equilibrium with the rest of the universe (thermal axions are subdominant for the mass range of interest).  In fact, axions are thought to form a Bose-Einstein Condensate~\citep{sikivie}.  In supersymmetric extensions of the Standard Model, axions would have massive fermionic partners, called axinos, which are also WIMP dark matter candidates.

The aforementioned candidates represent only a few of the myriad possibilities.  Given any theory of physics beyond the Standard Model, there will likely be at least one particle candidate for cold dark matter.  I have only discussed a few theories that naturally contain dark matter candidates, but many others exist.  For example, in Little Higgs models, dark matter could be heavy photons or scalars. Theories of Mirror Matter also contain dark matter candidates.  Or dark matter could be heavy sterile neutrinos. As many of these candidates are WIMPs, I restrict the discussion in the following sections to the status of direct and indirect searches for WIMP dark matter\footnote{For an overview of axion searches, please see~\citep{axionsearches}.}.

\section{Direct Detection}

\begin{figure}[h]
\includegraphics[height=.3\textheight]{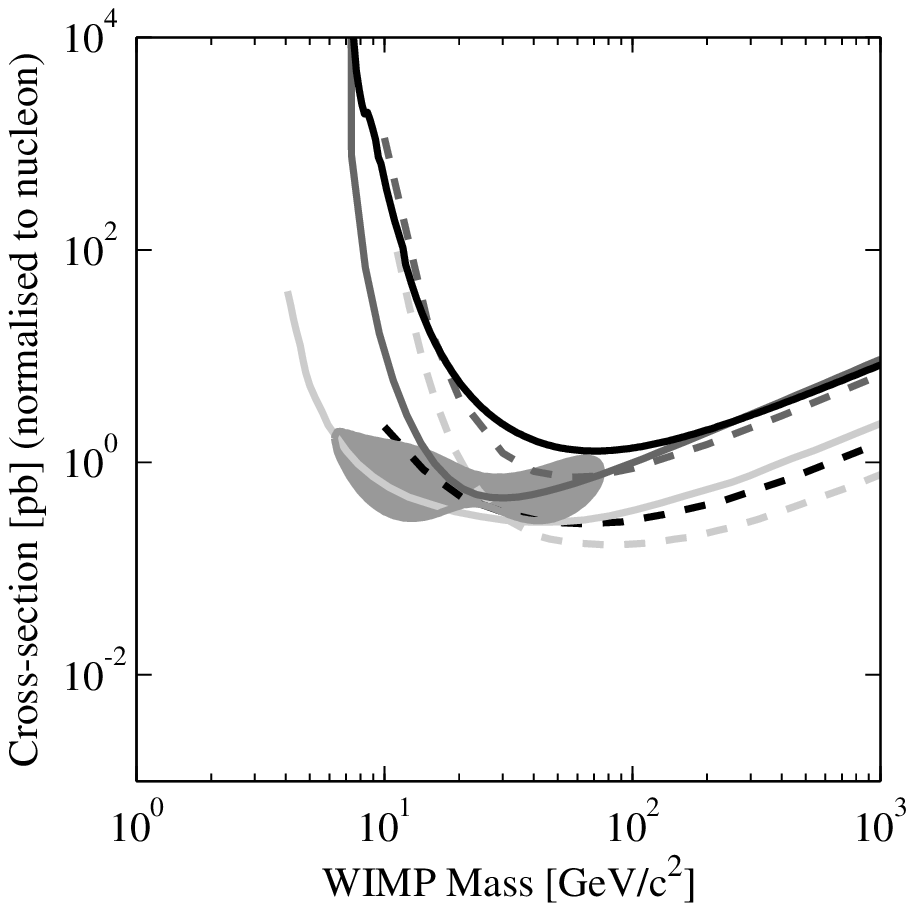}
\includegraphics[height=.3\textheight]{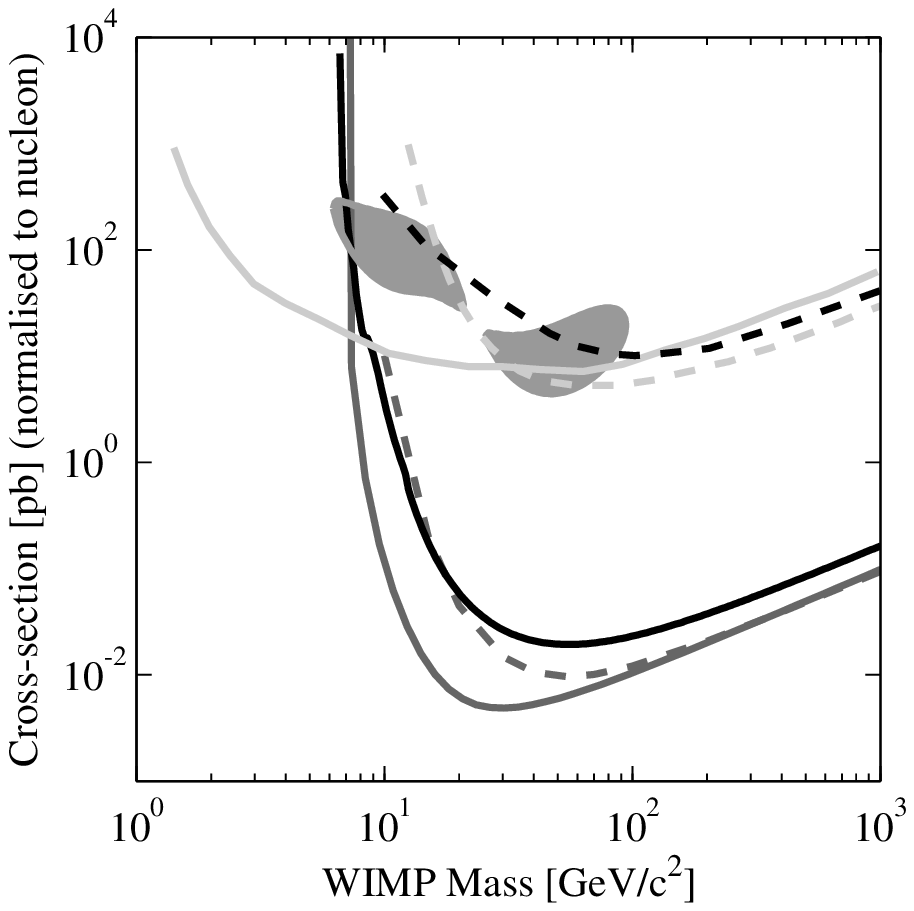}
\caption{In the left panel, we present the current limits on the spin-dependent WIMP-proton elastic scattering cross section from CDMS~\citep{cdms} (black solid), XENON10~\citep{xenonSD} (dark grey solid), ZEPLIN III~\citep{zep3SD} (dark grey dashed), NAIAD~\citep{naiadSD} (black dashed), COUPP~\citep{coupp} (light grey solid), and KIMS~\citep{kims} (light grey dashed). In the panel on the right, we show the current limits on the spin-dependent WIMP-neutron elastic scattering cross section for CDMS~\citep{cdms} (black solid), XENON10~\citep{xenonSD} (dark grey solid), ZEPLIN III~\citep{zep3SD} (dark grey dashed), KIMS~\citep{kims} (light grey dashed), NAIAD~\citep{naiadSD} (black dashed), and CRESST~\citep{cresstSD} (light grey solid). In each panel, we also show the region of the plane favored by the DAMA/LIBRA annual modulation signal at the $3\sigma$ level without ion channeling~\citep{dama} (medium grey shading). These plots, and that in Fig.~\ref{fig:directSI}, were created with the Dark Matter Tools interactive plotter~\citep{dmtools}.
\label{fig:directSD}}
\end{figure}

The most convincing verification of particle dark matter would be its direct observation through elastic scattering on nuclei. The scatterings are observed through the signatures of the recoiling nuclei, which can generally be resolved in a detector as phonons (heat), ionization, or scintillation.  There are many experiments that have used and/or are using these techniques to observe dark matter-nucleus scattering\footnote{Despite the use of the present tense, some of the experiments discussed in this section have been decommissioned.}. Single channel detectors are sensitive to only one byproduct of the WIMP-nucleus scattering, while two channel detectors measure the relative intensity of two effects, resulting in better background rejection. Among ionization-only detectors are experiments such as GENIUS, TEXONO, IGEX, HDMS, and CoGENT.  Experiments that look only for scintillation are DAMA, NAID, DEAP, CLEAN, XMASS, KIMS, and ANAIS. Cuoricino, CUORE, and CRESST-I are only sensitive to phonons, however CRESST-II is sensitive to both phonons and scintillation. EIDELWEISS, EUREKA, CDMS, and SuperCDMS are sensitive to both ionization and phonons, and liquid noble detectors such as ArDM, ZEPLIN, WARP, LUX and XENON are sensitive to both ionization and scintillation.  One may also look for evidence of the energy deposited in the detector by the scattering in a bubble chamber-type detector, as was done in PICASSO.  COUPP currently uses this technique to set the best limits on the spin-dependent (SD) WIMP-proton scattering cross section for WIMP masses $m_X \lesssim 30$ GeV~\citep{coupp}.  For $m_X \gtrsim 30$ GeV, KIMS~\citep{kims} has the best limit on this quantity\footnote{The limits on the SD WIMP-proton elastic scattering cross section from COUPP and KIMS are the best model-independent limits.  SuperKamiokande has performed an indirect dark matter search for neutrinos from annihilations in the Sun and can strongly constrain the SD WIMP-proton cross section for WIMPs annihilating dominantly to $b \bar{b}$~\citep{superk}. Indirect WIMP searches are discussed in Section~\ref{sec:indirect}}. In the case of WIMP-neutron scattering, XENON-10 has set the strongest constraint for $m_X \gtrsim 8$ GeV~\citep{xenonSD}, and CRESST for lower WIMP masses~\citep{cresstSD}. These results are displayed in Fig.~\ref{fig:directSD}.

\begin{figure}
\begin{center}
\includegraphics[height=.4\textheight]{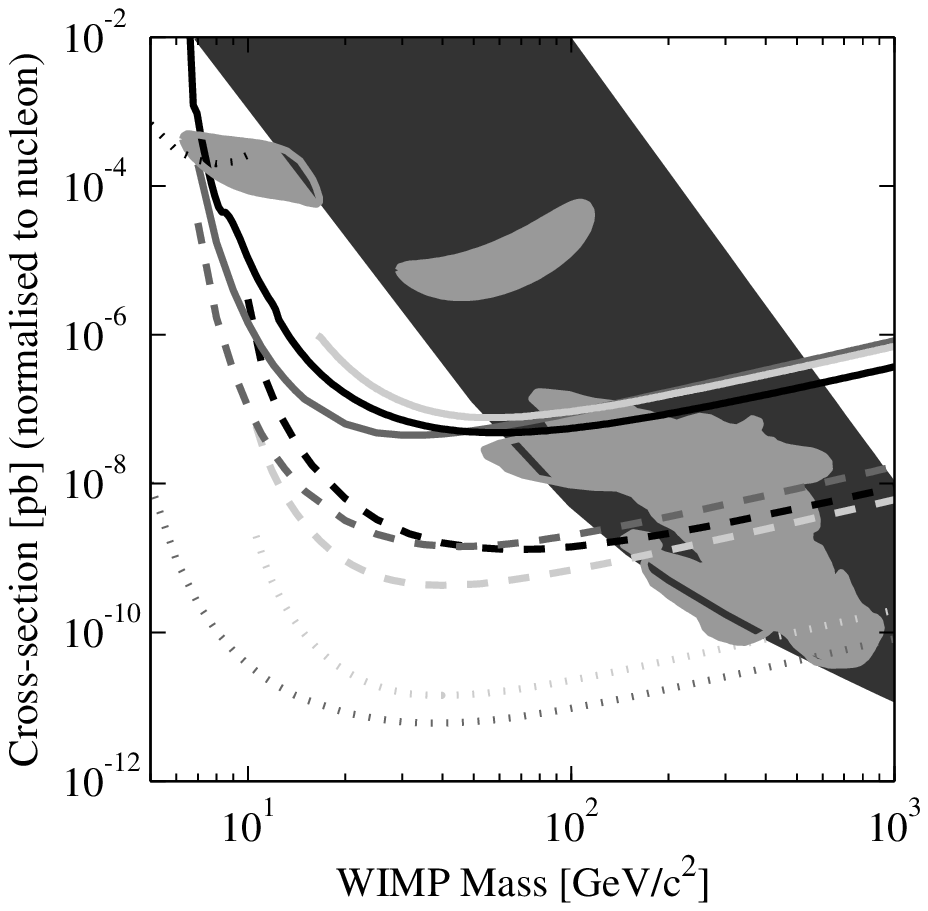}
\caption{Current limits on the spin-independent WIMP-nucleus elastic scattering cross section from CDMS~\citep{cdms} (black solid), and XENON10~\citep{xenon10} (dark grey solid), ZEPLIN III~\citep{zep3SI} (light grey solid), and CoGeNT~\citep{cogent} (black dotted, upper left corner). We also show the two regions favored by the DAMA/LIBRA annual modulation signal at the $3\sigma$ level without ion channeling~\citep{dama} (medium grey shading, $\sigma > 10^{-6}$ pb).  Dashed curves correspond to the projected sensitivities of SuperCDMS at Snolab (black), XENON100 (dark grey), and LUX 300 kg (light grey), and dotted curves to projections for future tonne-scale liquid noble detectors XENON1T (dark grey) and LUX/ZEP 3 Tonne (light grey)~\citep{dmtools}. The region of the plane favored for $B^{(1)}$ dark matter in Kaluza-Klein models with universal extra dimensions~\citep{arrenberg} (very dark grey shading), and that from minimal supersymmetric models with neutralino dark matter~\citep{trotta} (medium grey shading, $\sigma < 10^{-6}$ pb) are also shown.  
\label{fig:directSI}}
\end{center}
\end{figure}

In spin-independent (SI) WIMP-nucleus scattering, the whole nucleus participates coherently in the interaction, and one can increase the scattering cross section by building a detector of heavier target nuclei. As shown in Fig.~\ref{fig:directSI}, CDMS and XENON10 provide the best limits on the SI WIMP-nucleus elastic scattering cross section for $m_X \gtrsim 10$ GeV, and CoGeNT and TEXONO provide the best limits for very light WIMPs~\citep{cogent,texono}.

Theoretical predictions for the SI WIMP-nucleus scattering cross section can be calculated given the mass and couplings of a WIMP candidate.  The very dark grey shaded region in Fig.~\ref{fig:directSI} represents the projected cross sections for Kaluza-Klein dark matter, and the medium grey shaded region with $\sigma < 10^{-6}$ pb is that for neutralino dark matter.  Direct detection experiments are beginning to impinge on these regions.  The next generation of direct detection experiments, such as SuperCDMS and 100 kg-scale liquid noble detectors, with sensitivities shown as dashed lines, will do even better.  In the future, we can expect to cover nearly all of the scenarios shaded here, as liquid noble detectors will have sensitivities as low as $10^{11}-10^{12}$ picobarns, represented by the dotted contours in Fig.~\ref{fig:directSI}.  Note that the shaded regions represent predictions for only a few scenarios.  There are no guarantees that dark matter will be directly detected with current technology, but there seems to be good reason to have hope.

The direct detection experiments discussed up to this point search for individual dark matter scattering events by distinguishing nuclear recoils from electron scatterings and rejecting all backgrounds.  Another search strategy has been employed by the DAMA (and DAMA/LIBRA) Collaboration, which looks for an annual modulation signal of dark matter scatterings on top of the background. As the Sun moves through the Milky Way's dark matter halo, the Earth orbits the Sun, such that in June the Earth is moving with the Sun through the WIMPs and in December its motion is opposite the Sun's. We therefore expect an annual modulation of the WIMP-nucleus scattering rate, with an increased rate of scatterings in June and a decreased rate in December.

In fact, DAMA/LIBRA has observed an annual modulation signal at $8.2\sigma$ consistent with that expected from dark matter scattering on nuclei \citep{dama}. However, the result is in tension with the findings of other direct dark matter searches. In Figs.~\ref{fig:directSD} and~\ref{fig:directSI}, the region consistent with the annual modulation signal is shaded.  These DAMA-preferred regions are in portions of the parameter space already excluded by other experiments.

There are currently two types of suggested resolutions to the discrepancy: instrumental effects, such as ion channeling, or modified dark matter scattering kinematics, as in inelastic dark matter scenarios.  In the former case, channeled recoiling nuclei (ions) were shown to travel much farther in the detector than non-channeled ions~\citep{drobyshevski}.  In the case that ions are not channeled, recoiling nuclei lose their energy quickly by colliding with other nearby nuclei, thus depositing the bulk of the recoil energy in the detector as heat, which is not measured by DAMA/LIBRA.  If the ions are channeled, then they give energy, little by little, to electrons as they pass by.  In this case, nearly all the energy of the recoiling nucleus contributes to the detected scintillation signal.  The effect of ion channeling was examined by the DAMA Collaboration~\citep{bernabei} and others~\citep{savage}, and it was found that ion channeling increases the sensitivity to low WIMP masses, such that the DAMA region is disfavored by other experiments only at the $3\sigma$ level.

WIMP dark matter models in which the scattering kinematics are altered from the standard WIMP case are somewhat more successful at resolving the tension between DAMA and other direct searches~\citep{iDMhmm,iDMtsw}.  Inelastic dark matter scenarios involve WIMPs that have a slightly heavier excited state, $X^*$, where the mass splitting between the two states is $\delta = m_{X^*}-m_X$. In these scenarios, WIMPs scatter only inelastically off nuclei in the detector.  Scattering can only take place if the kinetic energy of the incoming WIMP is great enough to upscatter the WIMP into its excited state.
For a WIMP of mass $m_X$ and velocity $v_X$ colliding with a nucleus of mass $m_N$, the CM velocity is
\begin{equation}
v_{CM}=v_X\frac{m_X}{m_X+m_N},
\end{equation}  
and the required kinetic energy is 
\begin{equation}
\frac{1}{2} m_N v_{CM}^2 + \frac{1}{2} m_X (v_X-v_{CM})^2 > \delta c^2.
\end{equation}
One finds that 
\begin{equation}
\frac{1}{2}\Big(\frac{v_X}{c}\Big)^2\frac{m_Xm_N}{m_X+m_N} > \delta,
\end{equation}
so experiments with heavier nuclei are sensitive to larger mass splittings.  In fact, DAMA contains Iodine target nuclei, while CDMS consists of Germanium and Silicon, both significantly lighter. There is therefore a range of values of $\delta$ for which inelastic scattering of dark matter on nuclei would have been possible in DAMA/LIBRA but not in CDMS. Since Xenon is slightly heavier than Iodine, liquid Xenon detectors may soon confirm or refute the hypothesis of inelastic dark matter as the source of the annual modulation signal~\citep{XENONiDM}.

\section{Indirect Detection}
\label{sec:indirect}

In general, probes of astrophysical processes rely on our understanding of the elementary particles and their interactions.  At the same time, we can exploit the precision of these measurements to learn about particle physics itself.  Indirect detection of dark matter is one such enterprise.

In annihilations or decays of dark matter, all unstable particles will hadronize and/or decay to photons, electrons and positrons, neutrinos, and protons and antiprotons.  The spectra of these Standard Model particles produced in the annihilation or decay of dark matter depends on the dark matter candidate.  For example, supersymmetric neutralinos typically annihilate to $b\bar{b}$, $t \bar{t}$, or $W^+W^-$, with smaller branching fractions to other final states.  Quarks will immediately hadronize, and unstable states will decay to stable Standard Model particles, yielding different final state spectra in each case.  In this way, the observed spectra of annihilation products are related to the annihilation mode, and therefore to the theory of physics beyond the Standard Model in which the dark matter candidate is found.  

There are a plethora of experiments looking for these particles, in terrestrial, balloon-borne, and space-based detectors, at energies from sub-MeV to TeV and beyond.  Sites to search for indirect evidence of WIMP annihilations or decays include the core of the Sun~\citep{SOS}, the Earth~\citep{earth1,earth2}, our Galactic halo~\citep{zeldovich,rudaz,galaxy1,galaxy2,galaxy3,galaxy4}, Galactic center~\citep{GC}, and dwarf satellite galaxies~\citep{dsph1,dsph2}. Neutrinos from capture and scatterings in the Sun and Earth or from annihilation in the Milky Way or nearby dwarf satellite galaxies may be visible in Superkamiokande or in the IceCube neutrino detector at the South Pole. Ground-based Atmospheric \c{C}erenkov Telescopes, such as VERITAS, HESS, and MAGIC, and satellite-based telescopes EGRET and Fermi are sensitive to high energy photons that may be produced in annihilations in the Milky Way or in dwarf galaxies.  WMAP and Planck are sensitive to microwave radiation that may be due to synchrotron radiation of dark matter annihilation products in the magnetic fields near the Galactic center. And positrons and antiprotons from nearby annihilations in the Milky Way will be (and have been) probed by HEAT, ATIC, PAMELA, Fermi, and AMS.

In general, the differential flux of particles of type $j$ from the annihilation of dark matter particles of mass $m_X$ is
\begin{equation}
 \label{eq:dNdE}
 \frac{d \Phi_j(\Delta \Omega, E_j)}{dE_j} \propto 
        \frac{\langle \sigma v \rangle}{m_X^2}
 \sum_F f_F  \frac{dN_{j,F}}{dE_{j,F}} \times \mathcal{J},
\end{equation}
where $f_F$ is the fraction of annihilations to produce final state $F$, and \mbox{$dN_{j,F}/dE_{j,F}$} is the differential spectrum of particles $j$ from an annihilation to final state $F$. The calculation can be split into a part containing the particle physics properties of the dark matter and a part containing the information about the distribution of the source, $\mathcal{J}$. The source distribution enters the calculation as the square of the dark matter density, $\rho_X$, integrated along the line-of-sight, averaged over the solid angle $\Delta \Omega$, with some detector-dependent Point Spread Function, $PSF$:
\begin{equation}
 \label{eq:j}
 \mathcal{J} = \int_{\Delta \Omega} PSF\, d\Omega \,
         \int_{l.o.s.} \rho_X^2(s) ds.
\end{equation}

For charged annihilation products, diffusion and energy losses must be accounted for by solving the propagation equation,
\begin{equation}
\frac{\partial}{\partial t} \frac{dN}{dE} = \vec{\bigtriangledown}\cdot \left [ K(E,\vec{x})\,\vec{\bigtriangledown} \frac{dN}{dE} \right ] + \frac{\partial}{\partial E} \left [ b(E,\vec{x}) \,\frac{dN}{dE} \right ] + Q(E,\vec{x}),
\end{equation}
where $ K(E,\vec{x})$ is the diffusion coefficient, $b(E,\vec{x})$ is the rate of energy loss, and $Q(E,\vec{x})$ is the source term which reflects both the particle spectrum from annihilations and the dark matter source distribution~\citep{diffeq}.

Astrophysical processes typically produce much more matter than antimatter.  Dark matter, on the other hand, generally has no preference for matter over antimatter in its annihilations, so we expect equal numbers of electrons and positrons or protons and antiprotons from WIMP annihilation. Thus, the antimatter content of the cosmic ray spectrum may provide evidence of WIMP annihilations.  Notably, there is an unexplained excess of positrons (relative to electrons plus positrons) observed by PAMELA between 10 and 100 GeV~\citep{pamela}.  This excess was hinted at by HEAT~\citep{heat1,heat2} and AMS-01~\citep{ams01}, and is now well-measured up to 100 GeV.  ATIC showed a sharp peak in the spectrum of electrons plus positrons at $\sim500$ GeV~\citep{atic}.  The Fermi space telescope, which is also sensitive to a combined signal from electrons plus positrons, confirmed the excess in the few hundred GeV range, however the spectrum was measured to be significantly more flat~\citep{fermipos1,fermipos2}.  

Although the anomalies observed thus far could be attributed to more standard astronomical sources or a lack of understanding of the physical processes in our galaxy, it is exciting to consider the possibility that they might be due to the annihilations or decays of particle dark matter.  Interestingly, however, there is no observed excess of cosmic ray antiprotons over the expected background.  This indicates that if any of the positron signals are due to dark matter annihilation, dark matter must annihilate preferentially to leptons. Although some theories with this characteristic exist, so-called leptophilic annihilations present a significant model-building challenge.

Another model-building challenge comes from the annihilation rate itself.
The flux of dark matter annihilation products given by Eq.~\ref{eq:dNdE} is proportional to the annihilation cross section, $\langle \sigma v \rangle$, while, for particles that were initially in thermal equilibrium and whose abundance was fixed during freeze-out (in the manner of the WIMP Miracle), the relic abundance is inversely proportional to the annihilation cross section.    
Interestingly, for standard, thermally produced, weakly interacting dark matter, $\langle \sigma v \rangle \approx 3 \times 10^{-26}$ cm$^3$ s$^{-1}$, which leads to a flux of positrons insufficient to explain any of the current anomalies.
If one or more of the excesses in antimatter cosmic rays are due to dark matter annihilations, we are led to the conclusion that either the annihilation cross section today must be boosted relative to the cross section at freeze-out, or dark matter must have been produced in a non-thermal way, such that its annihilation cross section was always larger than the thermal expectation.

Such boosts to the annihilation cross section could have two potential origins.  In Eq.~\ref{eq:j}, we see that the flux of annihilation products is related to the dark matter density squared.  If the dark matter in the Milky Way halo is not smoothly distributed, but instead is clumpy, we would perceive it as a boost in the annihilation rate.  In fact, simulations suggest that the Milky Way halo is indeed quite clumpy~\citep{vl2}, however boosts from this type of structure only result in an increase in the flux by a factor of $\mathcal{O}(10)$~\cite{lavalle}.  A boost of $\mathcal{O}(10^2-10^3)$ is necessary to explain the PAMELA positron fraction for most WIMP dark matter.  For Kaluza-Klein dark matter, however, a boost of $\mathcal{O}(10)$ is sufficient.

A boost in the annihilation cross section could also be due to the particle theory itself.  For example, a larger-than-thermal annihilation cross section would be expected if annihilation proceeds via a Breit-Wigner resonance \citep{bweFeldman,bweIbe}.  Alternatively, if there is an attractive force between the annihilating dark matter particles mediated by a light gauge boson, the cross section could be significantly enhanced at low velocities. Since dark matter particles move much more slowly today, especially in substructure, than they were moving in the early universe, this so-called Sommerfeld enhancement~\citep{sommerfeld} could increase the annihilation cross section today by $\mathcal{O}(10^2)$ or more relative to the thermal expectation~\citep{cirelli,lattanzisilk}.  Of course, as mentioned above, a light force carrier is required.  For a WIMP with $m_X \approx 100$ GeV, a new gauge boson with mass $m_\phi \approx$ few GeV would be necessary.  If the dark matter is heavier, $m_X \gtrsim$ few TeV, then Standard Model gauge bosons could play the role of the light force carrier.  A Sommerfeld enhancement is therefore possible, for example, in supersymmetric theories with heavy Higgsino-like or wino-like neutralino dark matter~\citep{susySE}.

We now have several tantalizing signals of 100 GeV to TeV-scale positron excesses.  These excesses can be consistently fit with a variety of leptophilic particle dark matter candidates and with more ordinary astrophysical sources such as pulsars~\citep{fermipos2}.  An independent confirmation of the properties of particle dark matter is needed to break this degeneracy.

Neutral annihilation products, in contrast to the charged particles discussed above, have the advantage that they travel from the source to the observer without interacting significantly along the way.  Photons and neutrinos from dark matter annihilations have been and are currently being used to search for, among other things, dark matter annihilations and decays.  An excess of GeV photons was found by the EGRET satellite and had been interpreted as a signal of dark matter annihilations~\citep{deboer}.  But Fermi, currently in the process of performing a detailed all-sky survey in gamma-rays between 20 MeV and 300 GeV~\citep{fermigamma}, does not confirm the EGRET excess~\citep{porter}.  Fermi will eventually be sensitive to dark matter annihilations in many standard WIMP scenarios for cuspy dark matter halo profiles, although less so for less favorable halo profiles~\citep{fermigamma}.

Photons may come directly from the annihilation process and subsequent hadronization and decay of unstable particles, or they may come from synchroton radiation or inverse Compton scattering of charged annihilation products in the Galactic magnetic field.  In fact, there is an excess of microwave radiation towards the Galactic center evident in the WMAP data.  After subtraction of the known components (CMB, dust emission, soft synchrotron radiation from supernovae, and free-free emission), one finds that there is an additional component with a spectrum that is harder than that of synchrotron radiation from supernovae and incompatible with free-free emission.  Remarkably, with the very standard assumptions of WIMPs distributed according to an NFW halo profile annihilating with thermal cross section to a variety of standard model states in a $10 \mu$G magnetic field, it was found that the WMAP Haze was reproduced with roughly the right power, spectrum and spatial distribution~\citep{haze}.  Recently, there has been evidence of a corresponding Fermi Haze due to the same population of charged particles responsible for the WMAP Haze.  It has been speculated that the Fermi Haze, in the 1-100 GeV range, is due to upscattering of the interstellar radiation field by those charged particles~\citep{fermihaze}. Whether the origin is dark matter annihilation remains to be shown, but the possibility is intriguing, to say the least.

Like photons, neutrinos may also come directly from dark matter annihilations in the Milky Way or in nearby dwarf satellite galaxies.  Dwarf galaxies are promising sites to search for dark matter annihilation products because they are dark matter-dominated and contain few astrophysical sources that could mimic a signal.  Searches for neutrinos from dark matter annihilations are being performed at Superkamiokande and the IceCube neutrino detector, and will continue with the DeepCore addition to IceCube, which will increase its sensitivity to low energy neutrinos~\citep{deepcore}. KM3Net, to be constructed in the Mediterranean Sea, will also play an important role in the search for neutrinos from dark matter annihilation~\citep{km3net}.

Unlike other Standard Model particles, neutrinos are so weakly interacting that the Sun and Earth are not necessarily opaque to them.  If a WIMP scatters elastically with a nucleus in the Sun or the Earth, it may lose so much energy that the WIMP velocity falls below the escape velocity, at which point the WIMP is said to be ``captured.''  In this way, WIMPs accumulate in the core of the Sun or the Earth and annihilate.  The only annihilation products to reach terrestrial detectors are neutrinos, which could provide information about both the elastic scattering properties of the dark matter and the annihilation mode.

\section{Summary}

In this review talk I have briefly discussed particle candidates for dark matter arising in popular theories of physics beyond the Standard Model.  I have also reviewed the status of many direct and indirect searches for particle dark matter.  Indirect evidence for dark matter annihilations in the form of cosmic ray positron excesses, as well as potential gamma ray and microwave excesses, have generated considerable speculation about dark matter annihilations and decays. The variety of recent and expected astrophysical data, combined with the anticipated data from collisions at the LHC and the improved sensitivities of the next generation of dark matter direct detection experiments, make this a very interesting time for particle astrophysics.

\acknowledgements 

Pearl Sandick is supported by the National Science Foundation under Grant No.~PHY-0455649. 


\end{document}